\documentclass{aa}
\usepackage{amsmath}
\usepackage{natbib}
\usepackage{txfonts}
\usepackage{graphicx,multirow}
\usepackage{color}
\usepackage[]{hyperref}
\bibpunct{(}{)}{;}{a}{}{,}

\newcommand{\myemail}{bhusemann@aip.de}

\begin{document}

\titlerunning{PyCosmic: detecting cosmics in fiber-fed IFS datasets}
\title{PyCosmic: a robust method to detect cosmics in CALIFA and other fiber-fed integral-field spectroscopy datasets\thanks{\texttt{PyCosmic} is freely available as a \texttt{Python}-based stand-alone program at \url{http://pycosmic.sf.net} for download.}}
\author{B.\ Husemann\inst{1} \and S. Kamann\inst{1} \and C. Sandin\inst{1}  \and S. F. S\'anchez\inst{2,3} \and R. Garc{\'{\i}}a-Benito\inst{2} \and D. Mast\inst{2,3}}

\institute{Leibniz-Institut f\"ur Astrophysik Potsdam (AIP), An der Sternwarte 16, 14482 Potsdam, Germany, \email{\myemail}  
\and Instituto de Astrof\'isica de Andaluc\'ia (CSIC), Camino Bajo de Hu\'etor s/n Aptdo. 3004, E18080-Granada, Spain
\and
Centro Astron\'omico Hispano Alem\'an de Calar Alto (CSIC-MPIA), E-4004 Almer\'ia, Spain
}

\abstract{Detecting cosmic ray hits (cosmics) in fiber-fed integral-field spectroscopy (IFS) data of single exposures is a challenging task because of the complex signal recorded by IFS instruments. Existing detection algorithms are commonly found to be unreliable in the case of IFS data, and the optimal parameter settings are usually unknown apriori for a given dataset. }
{The Calar Alto legacy integral field area (CALIFA) survey generates hundreds of IFS datasets for which a reliable and robust detection algorithm for cosmics is required as an important part of the fully automatic CALIFA data reduction pipeline. Such a new algorithm needs to be  tested against the  performance of the commonly used algorithms \texttt{L.A.Cosmic} and \texttt{DCR}. General recommendations for the usage and optimal parameter settings of each algorithm have not yet been systematically studied for fiber-fed IFS datasets to guide users in their choice.}
{We developed a novel algorithm, \texttt{PyCosmic}, which combines the edge-detection algorithm of \texttt{L.A.Cosmic} with a point-spread function convolution scheme. We generated mock data to compute the efficiency of different algorithms for a wide range of characteristic fiber-fed IFS datasets using the Potsdam Multi-Aperture Spectrophotometer (PMAS) and the VIsible MultiObject Spectrograph (VIMOS) IFS instruments as representative cases.}
{\texttt{PyCosmic} is the only algorithm that achieves an acceptable detection performance for CALIFA data. We find that \texttt{PyCosmic} is the most robust tool with a detection rate of $\gtrsim90$\% and a false detection rate $\lesssim5$\% for any of the tested IFS data. It has one less free parameter than the \texttt{L.A.Cosmic} algorithm. Only for strongly undersampled IFS data does \texttt{L.A.Cosmic} exceed the performance of \texttt{PyCosmic} by a few per cent. \texttt{DCR} never reaches the efficiency of the other two algorithms and should only be used if computational speed is a concern. Thus, \texttt{PyCosmic} appears to be the most versatile cosmics detection algorithm for IFS data. It is implemented in the new CALIFA data reduction pipeline as well as in recent versions of the multi-instrument IFS pipeline \texttt{P3D}. Although \texttt{PyCosmic} has been optimized for IFS data, we have also successfully applied it to longslit data and anticipate that good results will be achieved with imaging data.}
{}
 
\keywords{Techniques: image processing - Methods: miscellaneous}
\maketitle

\section{Introduction}
The identification and rejection of artefacts on charged-couple device (CCD) detectors caused by cosmic ray hits (hereafter cosmics) is a persisting problem for the reduction and analysis of astronomical data. Combining multiple images of the same object or field is considered the best method to identify cosmics because it is less likely that the same pixel is affected in several images. Sophisticated algorithms that detect and reject outlier pixels during the combination of exposures were developed, for example, for the Hubble Space Telescope \citep[e.g.,][]{Fruchter:1997}. However, there are often cases where only a single exposure is available or multiple exposures cannot be combined. This happens frequently with fiber-fed integral field spectroscopic (IFS) data, where the effects of differential atmospheric refraction, instrument flexure, or a variable sky brightness during the sequence of exposures prevent reliable detection of cosmics by image comparison.

Various techniques have been developed to detect and reject cosmics in single CCD exposures. They use different methods like trained neural networks \citep{Salzberg:1995}, convolution with a point-spread function \citep[PSF,][]{Rhoads:2000}, Laplacian edge detection \citep[][hereafter D01]{Dokkum:2001}, image statistics \citep[][hereafter P04]{Pych:2004}, or a fuzzy logic approach \citep{Shamir:2005}. A detailed performance evaluation of the different algorithms on single astronomical images was presented by \citet{Farage:2005}. Their tests revealed that the D01 algorithm, also known as \texttt{L.A.Cosmic}, performed well on imaging data. The algorithm of P04, known as \texttt{DCR}, did not perform as well on images, but was much less computationally expensive and primarily designed for spectroscopic data. 

Currently, a thorough evaluation of the performance of cosmics detection algorithms for fiber-fed IFS data is missing. Signals in such data are much more complex because a spectrum from each individual fiber is recorded  along a discrete trace on the CCD, with little gaps between spectra.  Thus, edge-like structures are introduced, and bright object or night-sky emission lines are more likely to be misclassified as cosmics, which is why automatic data-reduction pipelines generally avoid including this crucial step in the reduction process  \citep[e.g.,][]{Barnsley:2012}. Sophisticated methods to detect cosmics in data from fiber-fed multi-object spectrographs were presented \citep{Zhu:2009,Wang:2009} and show excellent results, but their parameter choices seem arbitrary for the \texttt{L.A.Cosmic} and \texttt{DCR} algorithms as their prime reference. Additionally, there is no public code available to make an independent check of their results and to verify whether the algorithm works also with IFS data.

For the Calar Alto legacy integral field area (CALIFA) survey \citep{Sanchez:2012a} and other IFS studies using the same instrument \citep[e.g.,][]{Sandin:2008}, it was discovered that the available algorithms always selected night-sky or object-emission lines as cosmics for the IFS data. An initial attempt to reduce the high false detection rate for CALIFA data by using a simplified Laplacian edge detection algorithm was implemented into the \texttt{R3D} reduction package \citep{Sanchez:2006a} and was only partly successful. Although it reduced the number of false detections, a significant number of cosmics were undetected. 

In this paper, we present a novel algorithm called \texttt{PyCosmic}. It combines the iterative Laplacian edge detection scheme with a PSF convolution approach. We evaluate the performance of \texttt{PyCosmic} against the most popular algorithms available, \texttt{DCR} and \texttt{L.A.Cosmic}, on realistic mock data for different IFSs and compare the results with illustrative examples on observed raw data. We then provide general recommendations regarding the use of detection algorithms with fiber-fed IFS data.

The different algorithms used in this study are briefly described in Section~\ref{sec:methods}. Results of our detailed performance and parameter study on IFS mock data are then presented in Section~\ref{sect:parameter}, followed by results obtained for real data in Section~\ref{sect:real}. Finally, we provide general recommendations as a guide for other IFS users in Section~\ref{sect:guide}.

\section{Outline of different cosmics detection algorithms}\label{sec:methods}
In the following, we briefly describe the three algorithms used in our comparative study, including our novel \texttt{PyCosmic} algorithm, to understand their basic differences.
\subsection{\texttt{DCR}, count statistics on subframes}\label{sec:DCR}
A simple and fast algorithm was presented by P04, which uses count statistics to detect cosmics as outliers in the histogram of pixel counts. To do this, the image $I$ is first split into small overlapping subframes $I_i$ that are treated separately. These subframes are intentionally kept small, $\leq$100 pixels, to consider only a local distribution of counts. Thereafter, the mode $\langle I_i\rangle$ and standard deviation $\sigma_{I_i}$ are calculated for all pixels in a subframe. To remove the influence of high-value pixels, $\langle I_i\rangle$ and $\sigma_{i}$ are calculated a second time, this time only including pixel values $m$ that satisfy $\langle I_i\rangle-\xi\sigma_{i}<m<\langle I_i\rangle+\xi\sigma_{i}$, where $\xi$ is an arbitrary threshold value. 

The subsequent steps are: (i) construct a histogram $h(I_{i})$ using all pixel values $m$,  (ii) search for the first empty histogram bin with a value $m_0$ that is higher than $\langle I_i\rangle$, and (iii) find the first gap $[m_1,m_2]$ in the histogram with zero number counts that fulfils $(m_2-m_1)> \xi\sigma_{i}$ and $m_1>m_0$.

If such a gap exists, then all pixels in $I_i$ with a value higher than $m_1$ are masked as cosmics.
Masked pixels (including neighbor pixels inside a so-called ``growing radius'' of one to two pixels) are then replaced with the mean value of a set of nearby pixels. In most applications of \texttt{DCR}, the growing radius is set to one pixel to fully cover the boundaries of the cosmics, but we use a zero-growing radius to achieve a fair comparison with the results of other algorithms. Furthermore, to account for multiple pixel cosmics, the algorithm is run iteratively. There are three free parameters that have to be set: the shape of the subframes $I_i$, the threshold value $\xi$, and the number of iterations.

\subsection{\texttt{L.A.Cosmic}, The Laplacian edge detection approach}\label{sec:original}

D01 was the first to use the Laplacian operator for the detection of cosmics in astronomical images. In its discrete form, the Laplacian operator can be written as 
\begin{equation}
 \nabla^2f = \left\{\begin{matrix} 0 & -1 & 0\\ -1 & 4 & -1\\ 0 & -1 & 0\end{matrix}\right\}\quad .
\end{equation}
Convolved images using this operator will highlight sharp edges because it removes a smooth signal and increases the contrast of isolated strong pixels. 

The algorithm starts by subsampling  the bias-subtracted image $I$ by a factor $f_{\mathrm{s}}\!=\!2$. This subsampling is required to avoid attenuation of cosmics by negative cross patterns when convolving the image with the discrete Laplacian kernel
\begin{equation}
 \mathcal{L}^{(2)}=\nabla^2f\otimes I^{(2)}\quad ,
\end{equation}
 where $I^{(2)}$ is the subsampled image and $\otimes$ is the convolution operator. All negative values in $\mathcal{L}^{(2)}$  are set to zero before the image is resampled to its original size. We refer to the resulting image as $\mathcal{L}^{+}$.

Cosmics are identified in $\mathcal{L}^{+}$ with respect to the expected noise of each pixel. The noise properties of $\mathcal{L}^{+}$ and $I$ are nearly equal for higher standard deviations (D01), which is why $I$ can be used to estimate the noise ($N$),
\begin{equation}
N=\frac{1}{g}\left(g\left(M_5\otimes I\right)+\sigma_{\text{rn}}^2\right)^{\frac{1}{2}}\quad,\label{eq:noise}
\end{equation}
where $g$ is the gain [e$^{-}\text{ADU}^{-1}$], $\sigma_{\text{rn}}$ the readout noise [e$^{-}$], and $M_n$ an $n\!\times\!n$ median filter (here $n\!=\!5$ pixels). Deviations from the expected noise are calculated as
\begin{equation}
S=\mathcal{L}^{+}/\left(f_{\text{s}}N\right)\quad.
\end{equation}

\begin{figure*}
 \includegraphics[width=\textwidth]{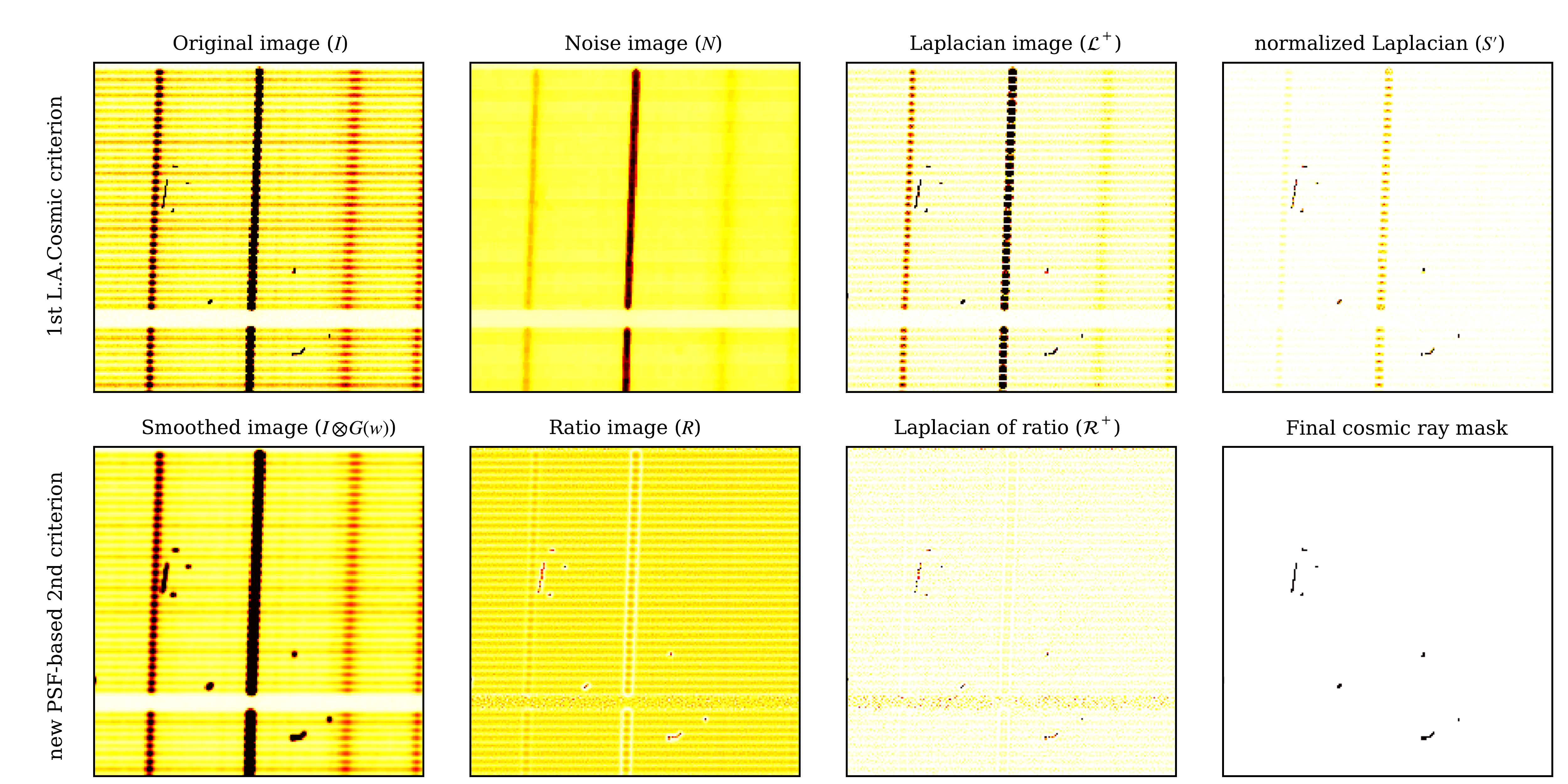}
\caption{Visual outline of the intermediate steps of our \texttt{PyCosmic} algorithm, which is based on \texttt{L.A.Cosmic}.}
\label{fig:process}
\end{figure*}

Signals of real objects remain in $S$ because of Poisson noise and the pixel sampling of smooth intensity profiles (cf. Fig.~\ref{fig:process}). This component, the sampling flux, can be significant if the signal is high or if the PSF is poorly sampled. Extended structures, which are larger than about five pixels, can be removed from $S$ using a $5\!\times\!5$ median filter; $S'=S-M_5\otimes S$. The first criterion for detection of cosmics demands $S'>\sigma_{\text{lim}}$, where a typical limiting value is $\sigma_{\text{lim}}=5$.

In addition, it will be difficult to distinguish cosmics from stars in a critically sampled image, i.e., close to Nyquist sampling, because they are very similar on small scales. Such point sources can, however, be distinguished from cosmics by their symmetry. An image $\mathcal{F}$ is calculated that contains only symmetric fine structure on scales of 2--3 pixels
\begin{equation}
\mathcal{F}=M_3\otimes I-\left[M_7\otimes\left(M_3\otimes I\right)\right].
\end{equation}
The second criterion states that the contrast between $\mathcal{L}^{+}$ and $\mathcal{F}$ is greater than a limiting value, $\mathcal{L}^{+}/\mathcal{F}>f_{\text{lim}}$, where typical limiting values for images are $f_{\text{lim}}\!=$2.0--5.0. 

Cosmics are finally identified as pixels that satisfy both criteria, although cosmics are mostly larger than a single pixel. While detection probability is higher for pixels on the edge of large multiple-pixel features, it may be negligible for pixels within the feature. Arbitrarily large cosmics can be fully detected by iteratively applying the rejection process as described above. After each iteration, the newly identified cosmics are replaced with the median of nearby unmasked pixels. In total, there are four free parameters to set: the threshold value $\sigma_\mathrm{lim}$, the limiting value $f_\mathrm{lim}$, the number of iterations, and a special parameter $\sigma_\mathrm{frac}$. 
This last undocumented parameter is a factor that is used to reduce the $\sigma_\mathrm{lim}$ threshold of neighbor pixels within a growing radius to detect cosmics in them as well. By default, the growing radius is set to one pixel. By definition, the effective growing radius is set to zero when $\sigma_\mathrm{frac}=1.0$.

\subsection{\texttt{PyCosmic}, combining edge detection with a PSF convolution approach}\label{sec:new}

Although \texttt{L.A.Cosmic} performs extremely well on imaging data, it is much less effective with fiber-fed IFS data. Spectra of several hundred fibers are dispersed along one axis of the CCD and are closely packed on the other axis with small separation. This introduces a highly asymmetric situation on small pixel scales, which is why neither of the \texttt{L.A.Cosmic} criteria are robust in this case. The longslit spectroscopy version of \texttt{L.A.Cosmic} invokes a model fit to the sky and object spectra. However, this scheme is difficult to apply to fiber-fed IFS data due to the comparatively inhomogeneous distribution of spectra on the CCD. It is practically impossible to fit thousands of profiles without introducing additional edges.

Our novel approach replaces the second criterion of \texttt{L.A.Cosmic} to avoid the simple median smoothing. Instead, we take advantage of the smooth two-dimensional shape of the spectrograph PSF in contrast to the highly asymmetric cosmics. This combines PSF-matched filtering \citep{Rhoads:2000} and Laplacian edge detection of cosmics in a simple and effective way. 

To discriminate between real signal and cosmics, we first smooth the bias-subtracted image $I$ by convolving it with a two-dimensional Gaussian kernel $G(w)$  (where $w$ is the full width at half maximum (FWHM) of the Gaussian in pixels) and divide $I$ by the smoothed image,
\begin{equation}
  R = \frac{I}{I \otimes G(w)}\quad .
\end{equation}
The idea is to increase the contrast of higher frequency signal compared to the object signal, so that $w$ should be \emph{larger} than the width of the cosmics and \emph{smaller} than FWHM of the spectrograph PSF ($\theta$). We note that a similar approach was independently used by \citet{Conselice:2003} to define the clumpiness parameter as a measure for the high-frequency components in the morphology of galaxies. The artificial smoothing of the data is a computationally easy task and the most natural choice to capture the high-frequency components in a signal.
Cosmics with $R\gg1$ appear surrounded by a halo with values of $R\ll 1$. When $R\approx1$, however, there is a homogeneous structure. We further increase the contrast of cosmics in $R$  by convolving it with the Laplacian kernel after subsampling $R$ by a factor of two,
\begin{equation}
 \mathcal{R}^{(2)}=\nabla^2f\otimes R^{(2)}\quad .
\end{equation}
Again, negative values in $\mathcal{R}^{(2)}$ are set to 0, and the image is re-sampled to its original size; we refer to this result as $\mathcal{R}^+$. 

We then replace the second criterion of \texttt{L.A.Cosmic} by $\mathcal{R}^+>r_\mathrm{lim}$, in order to minimize the false detection of real signal in fiber-fed IFS data. Snapshots of intermediate images of the process steps are shown in Fig.~\ref{fig:process}, which give a visual impression of the algorithm. 

Before the Laplacian convolution is applied to the image in each subsequent iteration, it is necessary to replace all the pixels of the detected cosmics with ``good'' values. Yet, it is very difficult to restore the original information of these pixel in IFS data, particularly near bright emission lines. An artificial edge could be created that can cause the cosmics to expand into the unaffected signal of the line. However, it is straightforward to mask all pixels of already detected cosmics in the convolution operation $I \otimes G(w)$. In this way, we minimize the artificial extension of cosmics into bright object data. \texttt{PyCosmic} does not employ the $\sigma_\mathrm{frac}$ parameter and thus has one less free parameter than \texttt{L.A.Cosmic}.

\begin{figure}
\resizebox{\hsize}{!}{\includegraphics{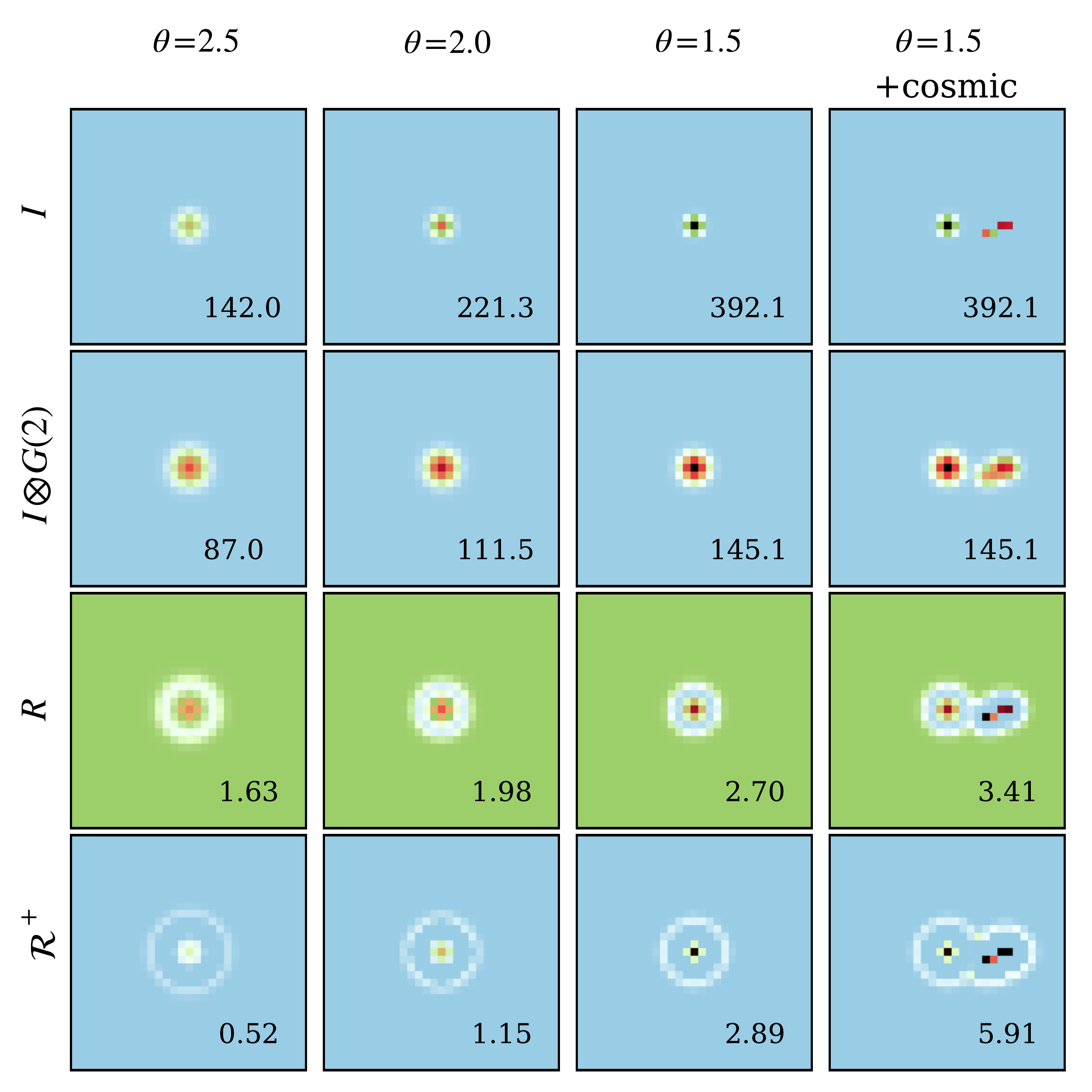}}
 \caption{Simulated images to estimate the minimum threshold value for $r_\mathrm{lim}$. Simulated thumbnail images and intermediate images to compute $\mathcal{R}^+$ are shown for a pure emission line feature considering three different instrumental PSFs with $\theta=2.5$, $2.0$, and $1.5$\,pixels FWHM. The maximum pixel value is provided for each subimage. The maximum in $\mathcal{R}^+$ corresponds to the minimum value for $r_\mathrm{lim}$ to avoid misclassification as cosmics. }
 \label{fig:flim_images}
\end{figure}

\begin{figure} 
\resizebox{\hsize}{!}{\includegraphics{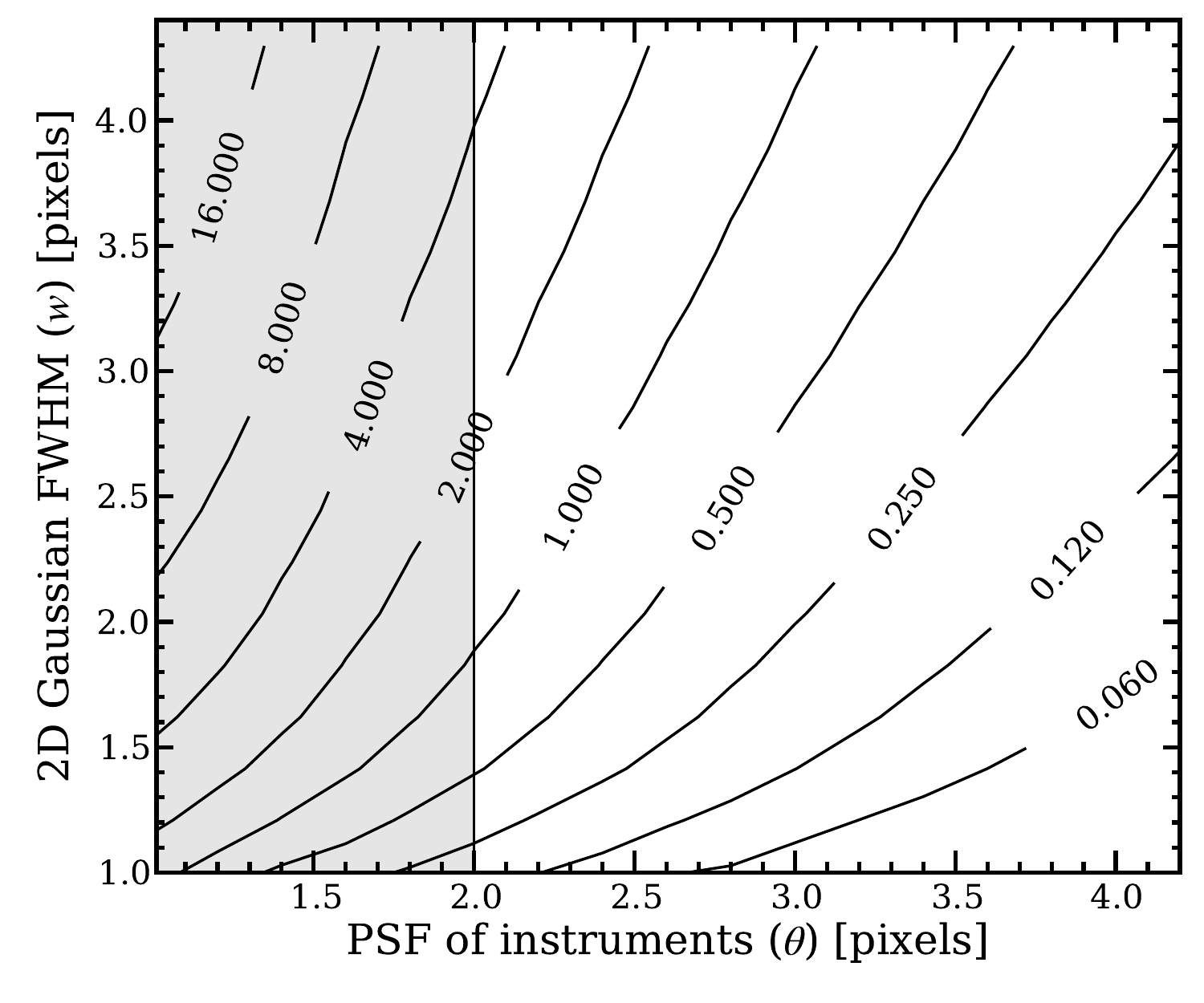}}
  \caption{Absolute minimum values of $r_\mathrm{lim}$ as a function of $\theta$ and $w$. These values were determined from noise-free simulations and set a hard lower limit to avoid frequent false detections of object signal as cosmics. They are estimated from the instrument-specific value of $w$. The shaded area highlights the regime of significantly undersampled data, where more care is needed.}
 \label{fig:flim}
\end{figure}

The threshold value $r_\mathrm{lim}$ depends on two independent parameters: the FWHM $w$ of the Gaussian kernel $G(w)$ and the FWHM of the instrumental PSF $(\theta)$.  We consider here that $\theta$ and $w$ are both identical for the dispersion and the cross-dispersion directions on the CCD. This is a realistic assumption for most IFS instruments, except when pixels are binned differently on the x- and the y-axes during CCD read-out. In that case, the axis with the smallest $\theta$ value is used as reference to select $r_\mathrm{lim}$ with respect to $w$ of the round Gaussian kernel. To estimate a minimum $r_\mathrm{lim}$ value for a given setup, we simulated and processed snapshot images of a single emission line for a grid of $\theta$ and $w$ values. In Fig.~\ref{fig:flim_images} we show the $\mathcal{R}^+$ image for a few simulated cases with $1.5\leq\theta\leq2.5$ and $w=2$.  The maximum value in $\mathcal{R}^+$ defines the absolute minimum value of $r_\mathrm{lim}$ to avoid misidentification of object signal as cosmics. The results from the parameter study with $1.0\leq\theta\leq4.2$ and $1.0\leq w\leq4.5$ are summarized in Fig.~\ref{fig:flim}, which serves as a guideline for selecting an appropriate $r_\mathrm{lim}$ for any dataset. However, these are idealized values in the sense that no noise, no underlying continuum, and no cross-talk between the different fibers have been taken into account. Hence, the optimal $r_\mathrm{lim}$ threshold for real data should be slightly higher.

\section{Performance tests of the detection algorithms}\label{sect:parameter}
\subsection{Simulation of mock fiber-fed IFS data}
\begin{table}
\begin{footnotesize}
\caption{Detection quality of the best parameters for the simulated data}
\label{tab:parameter}
\begin{tabular}{lccc}
  \hline\hline\noalign{\smallskip}& PMAS-LArr & PMAS-PPak & VIMOS MR\\\hline\noalign{\smallskip}
CCD size & 4kx4k & 4kx4k & 2kx4k\\
CCD binning & 2x2 & 2x2 & 1x1\\
grism & R1200(bw) & V500 & MR\\
$\theta$ & 1.5 & 2.4 & 3.0\\
$N_c$ & 6386 & 6471 & 11726 \\\hline\noalign{\smallskip}
\multicolumn{4}{c}{\texttt{DCR}}\\\hline\noalign{\smallskip}
$P_\mathrm{d}$ [\%] &  65.6 & 74.6 & 83.6 \\
$P_\mathrm{f}$ [\%] &  6.2 & 3.1 & 0.5 \\
$\epsilon$ [\%]& 62.9 & 73.0 & 83.1 \\
$I_i$ & 20$\times$20 & 10$\times$10 & 5$\times$20\\
$\xi$ & 7.0 & 3.0 & 3.0 \\\hline\noalign{\smallskip}
\multicolumn{4}{c}{\texttt{L.A.Cosmic}}\\\hline\noalign{\smallskip}
$P_\mathrm{d}$ [\%] &  95.2 & 86.9 & 94.5 \\ 
$P_\mathrm{f}$ [\%] &   3.2 & 39.2 & 3.3  \\ 
$\epsilon$ [\%] &  \textbf{92.8} & 61.4 & 91.5 \\ 
$\sigma_\mathrm{frac}$ & 1.0 & 1.0 & 1.0 \\
$f_\mathrm{lim}$ & 23 & 27 & 7 \\\hline\noalign{\smallskip}
\multicolumn{4}{c}{\texttt{PyCosmic}}\\\hline\noalign{\smallskip}
$P_\mathrm{d}$ [\%]      &  92.9 & 92.5 & 96.1 \\
$P_\mathrm{f}$ [\%]      &  5.5 & 1.3 & 3.5\\
$\epsilon$ [\%] &  88.7 & \textbf{91.6} & \textbf{92.9} \\
$w$ & 1.0 & 2.0 & 1.5 \\
$r_\mathrm{lim}$ & 1.25 & 1.4 & 1.0 \\\hline
\end{tabular}
\end{footnotesize}
\end{table}
We prepared dedicated simulations to test the performance of \texttt{PyCosmic} against \texttt{DCR} and \texttt{L.A.Cosmic}. In order to obtain unbiased results from the simulations, it is important to ensure that the signal distribution and shapes of cosmics are as realistic as possible. As we mentioned earlier, the $\sigma$-clipping is unfeasible for the majority of IFS data. Instead, we use dark frames to extract our template cosmics masks for two telescopes/instruments: the Potsdam Multi-Aperture Spectrophotometer \citep[PMAS,][]{Roth:2005} at the 3.5m Calar Alto telescope and the VIsible MultiObject Spectrograph \citep[VIMOS,][]{LeFevre:2003} at the Very Large Telescope. Dark frames are ideally suited for our purpose because they do not contain any signal, yet their long exposure times ($\sim$\,1800\,s) are comparable to those of typical science frames. We determined the noise level in the dark frames and selected all outliers above a $5\sigma$-threshold as cosmics. Given the low dark current and read-out noise of the detectors, the $5\sigma$ limit corresponds to $\sim$30 counts, the minimum signal of recovered cosmics.

The PMAS instrument offers two integral-field units (IFUs): a lens array (LArr) with $16\times16$ lenses and a simple bundle of 382 fibers \citep[PPak,][]{Kelz:2006}, with the latter used in the CALIFA survey \citep{Sanchez:2012a}. VIMOS is a versatile imaging and multi-object spectrograph, which  also includes a lens-array IFU.  We simulated IFS raw data including the night-sky spectra as the main signal, which is understood well from existing observations for the three following IFU instruments/setups: 
\begin{itemize}
\item[1)] PMAS-LArr IFU with a R1200 grating backward (bw) mounted and a 2x2 CCD binning, leading to $\theta\sim1.5$\,pixels,
\item[2)] PMAS-PPak IFU with a V500 grating and a 2x2 CCD binning resulting in $\theta\sim2.4$\,pixels (CALIFA survey setup),
\item[3)] VIMOS IFU with the mid resolution (MR) grism operating at $\theta\sim3$\,pixels.
\end{itemize}
These IFU setups cover a wide range in data sampling from significantly undersampled to very well sampled data. We consider them representative of all fiber-fed IFUs that are currently in operation.
Synthetic IFS data for each setup were produced  using reduced fiberflats, traces, and dispersion masks from real observations. The same observed night-sky spectrum, scaled to 1800\,s effective exposure time, was used as input signal for all fibers. Afterwards, Possion and read-out noise were added to the simulated images, as well as empirical cosmics from the dark frames.

\subsection{Parameter study to reach optimal performance}
We did not include additional signal from astronomical objects, given that the simulated IFU data of the night sky already include continuum and bright emission line features similar to the characteristics of astronomical objects. The goal here was to test the performance of the detection algorithms when the signal in each spectrum is dominated by the sky rather than the typically fainter object signal.

In order to properly compare the performances of each algorithm, we tested them with a grid of input parameters because the optimal ones are unknown apriori for a given dataset. From the simulations, we defined the number of detectable cosmics $N_\mathrm{c}$, which were $5\sigma$ above the noise of the simulated image before the cosmics were added. We defined the detection rate as $P_\mathrm{d}=N_\mathrm{d}/N_\mathrm{c}$, with the number of detected cosmics that match the input mask ($N_\mathrm{d}$). Pixels that were misclassified as cosmics by the algorithms are false detections ($N_\mathrm{f}$), expressed as a false detection rate $P_\mathrm{f}=N_\mathrm{f}/N_\mathrm{c}$. We defined the detection efficiency as $\epsilon=N_\mathrm{d}/(N_\mathrm{c}+N_\mathrm{f})$, which has a value of $\epsilon=1$ for ideal detection rates and $0<\epsilon<1$ when the detection was incomplete or the number of false detection was non-zero.

\begin{figure*}
\centering
 \includegraphics[width=0.9\textwidth]{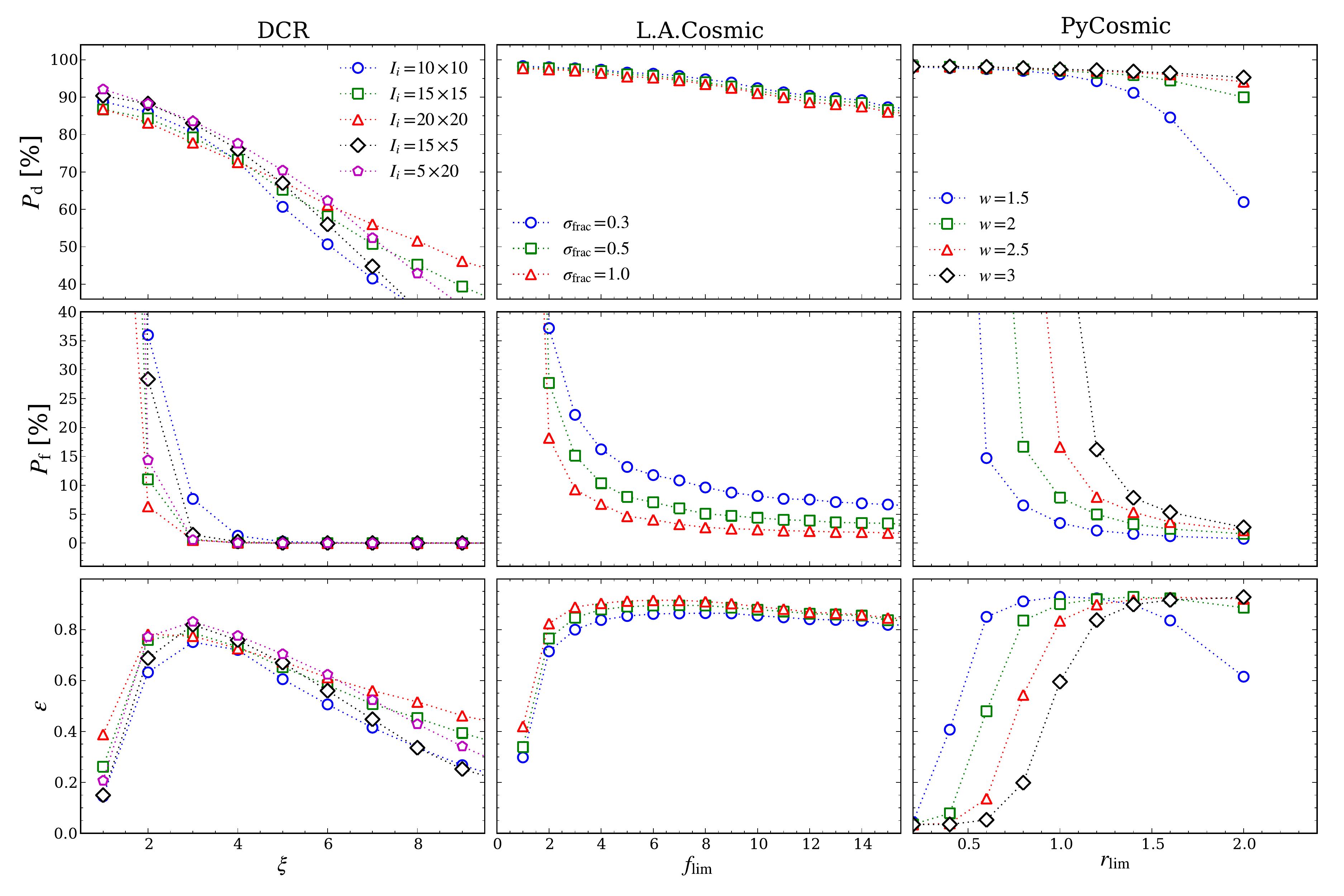}
  \caption{Performance and parameter study of the \texttt{DCR}, the \texttt{L.A.Cosmic}, and the \texttt{PyCosmic} algorithms applied to simulated VIMOS IFU data in the MR mode with $\theta=3$\,pixels. The first row of all panels compares the detection rate ($P_\mathrm{d}$) of cosmics, the second row compares the false detection rate ($P_\mathrm{f}$), and the third row shows the combined efficiency ($\epsilon$). The primary parameter that controls the performance of the algorithms is varied along the abscissa of each column of panels, while a secondary parameter is shown as different symbols connected by dotted lines as assigned in the legend.}
 \label{fig:parameter1}
\end{figure*}

Each of the algorithms has free parameters that need to be chosen by the user: 
\begin{itemize}
 \item[a)] \texttt{DCR:} subframe size of $I_i$, limiting sigma factor $\xi$, the number of iterations, and growing radius (set to zero pixels),
 \item[b)] \texttt{L.A.Cosmic}: significance $\sigma_\mathrm{lim}$, threshold $f_\mathrm{lim}$, threshold $\sigma_\mathrm{frac}$, and the number of iterations
 \item[c)] \texttt{PyCosmic}: significance $\sigma_\mathrm{lim}$, threshold $r_\mathrm{lim}$ appropriate for the chosen value of $w$ and the instrument specific value of $\theta$, and the number of iterations.
\end{itemize}
We consistently used a maximum number of six iterations in all cases for our tests. For \texttt{L.A.Cosmic} and \texttt{PyCosmic}, we set the significance level to $\sigma_\mathrm{lim}=5$ to achieve comparable results. The algorithm performance as a function of input parameters is summarized in Figs.~\ref{fig:parameter1}-\ref{fig:parameter3} for the three IFS instruments. Surprisingly, we found that the achievable performance was strongly dependent on the IFS instrument characteristics.

For the simulated VIMOS IFU data, which is representative of well-sampled raw data with $\theta\sim3$\,pixels, the \texttt{L.A.Cosmic} and \texttt{PyCosmic} algorithms performed almost equally well at their best-parameter settings with a detection rate of $P_\mathrm{d}\sim95\%$. \texttt{PyCosmic} achieved a slightly higher detection ($P_d=96.1\%$) with a marginally higher false detection rate ($P_f=3.5\%$) compared to \texttt{L.A.Cosmic} ($P_d=94.5\%$, $P_f=3.3\%$). The detection rate of $P_\mathrm{d}=80\%$ for \texttt{DCR} may not be sufficient for many applications.

Interestingly, \texttt{L.A.Cosmic} showed the poorest performance for PMAS-PPak IFU data critically sampled at $\theta\lesssim2.4$\,pixels. The instrumental characteristics responsible for this substandard performance remain unclear. All parameter configurations gave $P_\mathrm{f}\gtrsim 40$\%, which is unacceptable. \texttt{PyCosmic} was clearly the best algorithm, with a high detection rate and an accompanied low false detection rate ($P_\mathrm{f}<1.5\%$). \texttt{DCR} performed as poorly as \texttt{L.A.Cosmic} with low detection and high false detection rates.

Simulated data for the highly undersampled PMAS-LArr setup ($\theta\sim1.5$\,pixels) are clearly domains of \texttt{L.A.Cosmic}, because it was initially optimized for strongly undersampled Wide-Field Plenetary Camera 2 Hubble Space Telescope images. \texttt{L.A.Cosmic} reached a  high detection rate at an acceptable false detection rate. Nevertheless, \texttt{PyCosmic} achieved  a similar efficiency when the smoothing kernel width was set to $w=1.0$\,pixels, significantly smaller than the instrumental PSF. We expected \texttt{DCR} to perform poorly on undersampled IFU data because real signal and cosmics are hard to distinguish from count statistics. This is confirmed by our simulations. 

\begin{figure*}
 \includegraphics[width=\textwidth]{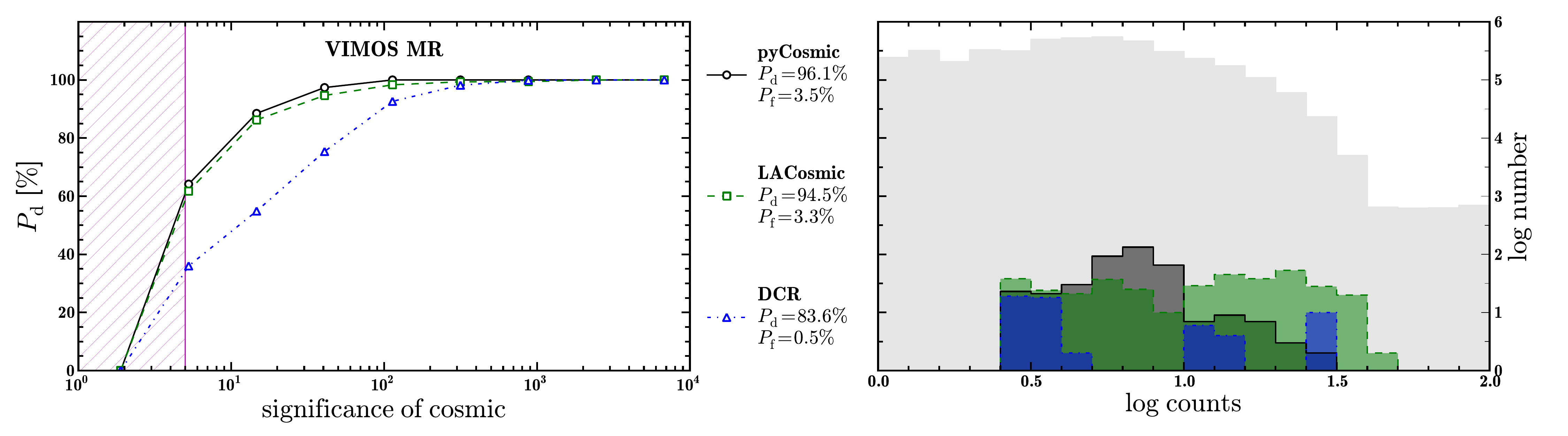}\\
  \includegraphics[width=\textwidth]{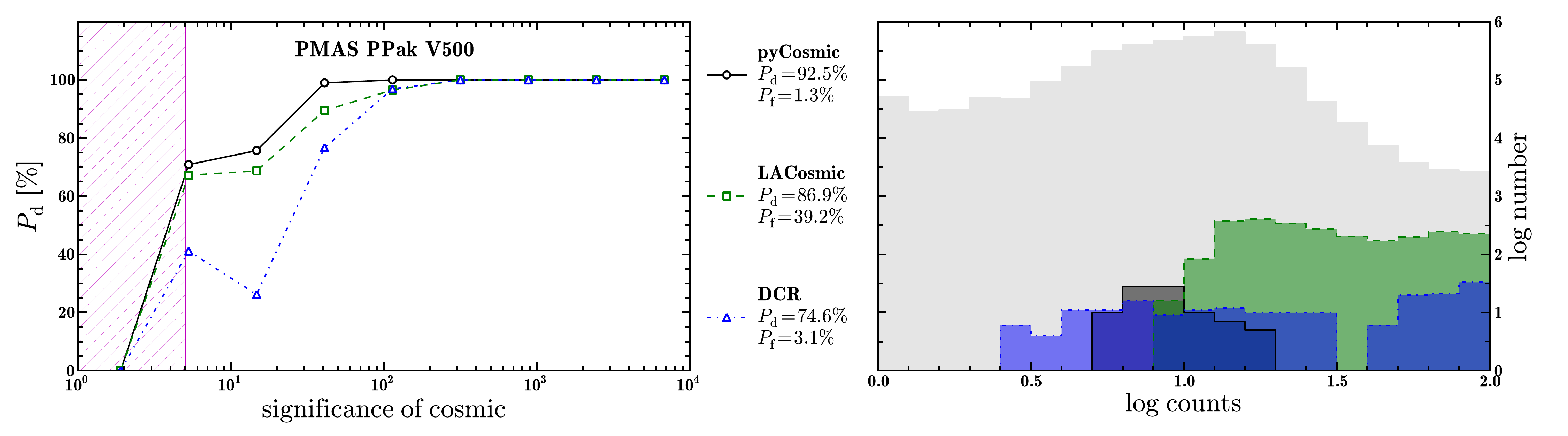}\\
 \includegraphics[width=\textwidth]{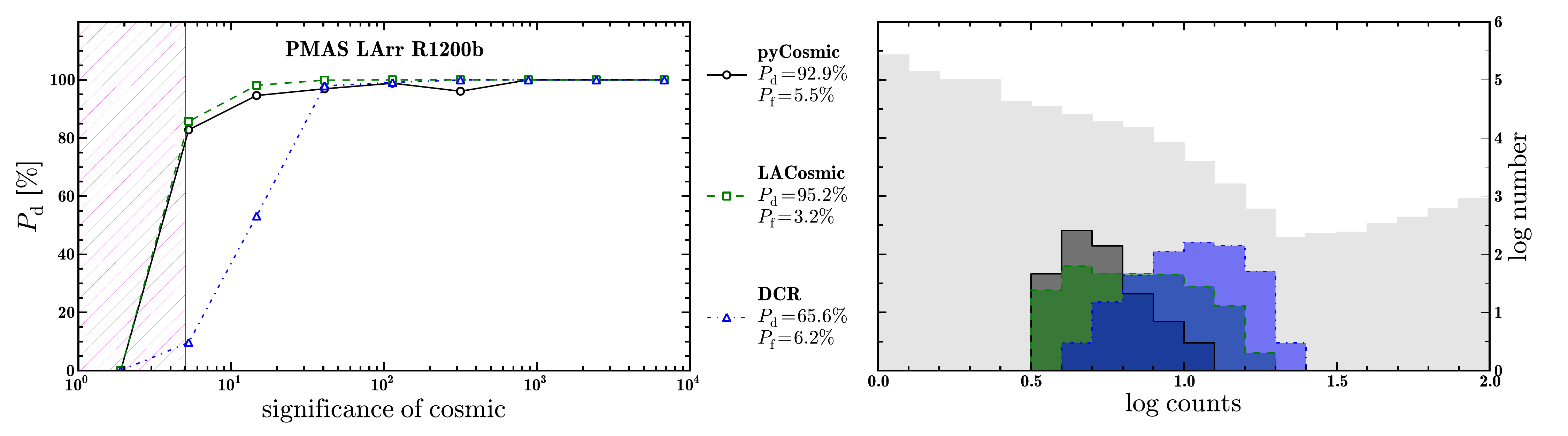}
 \caption{Detailed statistics of the detection rate and spuriously detected pixels for the three simulated raw datasets (\texttt{PyCosmic}$\rightarrow$black, \texttt{L.A.Cosmic}$\rightarrow$green, and \texttt{DCR}$\rightarrow$blue). \textit{Left panels:} Detection rate in per cent as a function of the cosmic ray significance, i.e., cosmics counts divided by the Poisson and the read-out noise of the simulated pixel. \textit{Right panels:} Histogram of pixel counts for false detection compared to the count histogram of the entire image (grey area). The counts are given in logarithmic units. }
 \label{fig:simulations}

\end{figure*}

Table~1 summarizes the algorithm parameters together with their corresponding $P_\mathrm{d}$ and $P_\mathrm{f}$ rates for each setup that gives the highest efficiency. For these ``optimal'' parameter settings, we made an additional statistical comparison to further elucidate strengths and weaknesses of each algorithm. The results are shown in Fig.~\ref{fig:simulations}. \texttt{L.A.Cosmic} and \texttt{PyCosmic} behaved similarly well as $P_\mathrm{d}$ steeply rose to $P_\mathrm{d}=100\%$ above the threshold significance of $5\sigma$. The \texttt{DCR} algorithm in all cases had a shallower curve and reached 100\% efficiency at much higher significance than the other two. Concerning misclassified pixels, we found that \texttt{PyCosmic} tended to detect pixels with low counts as cosmics, while \texttt{L.A.Cosmic} did not. 

\section{Illustrative examples of cosmics detection for observed IFS data}\label{sect:real}
\begin{figure}
\resizebox{\hsize}{!}{\includegraphics{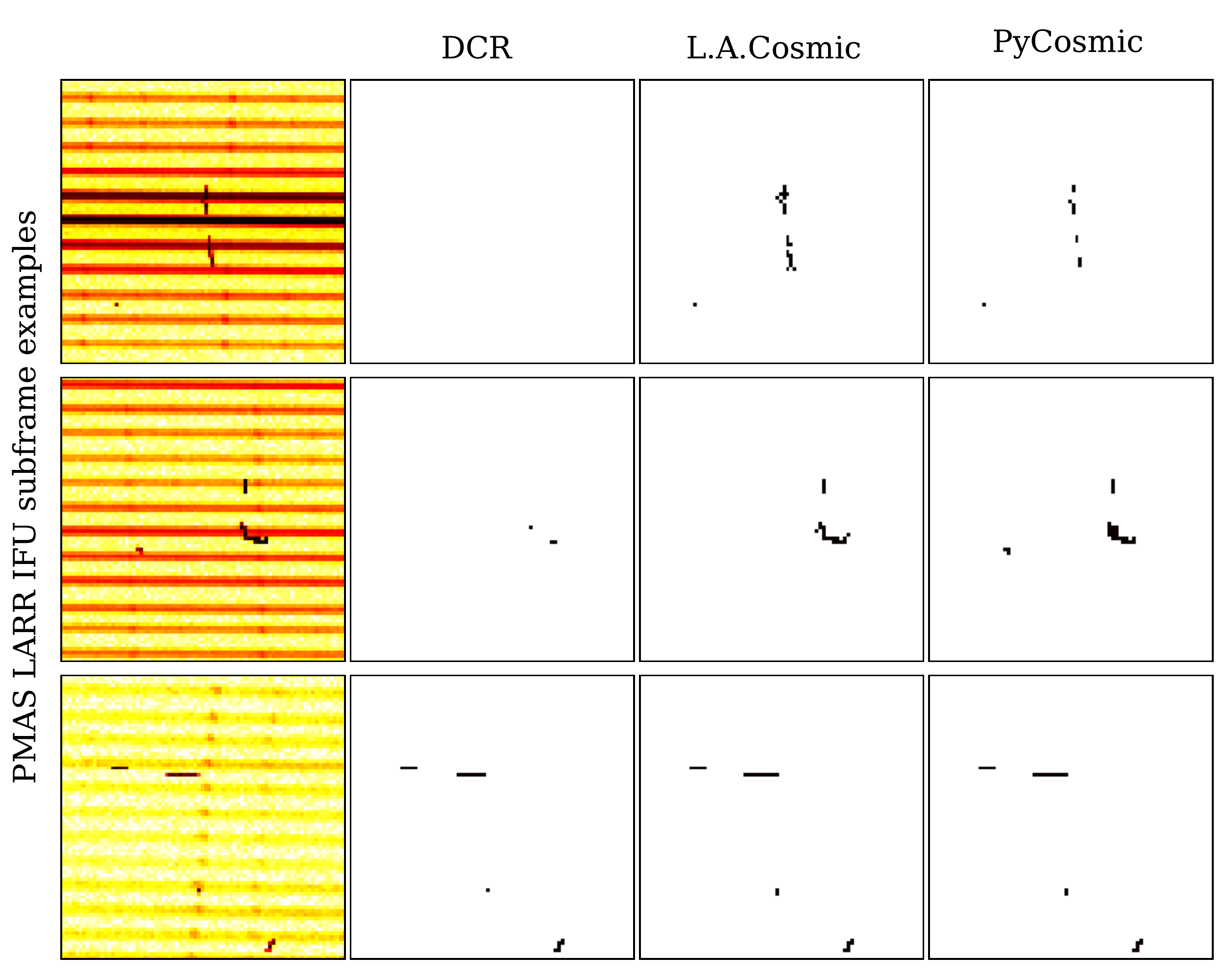}}\\
\resizebox{\hsize}{!}{\includegraphics{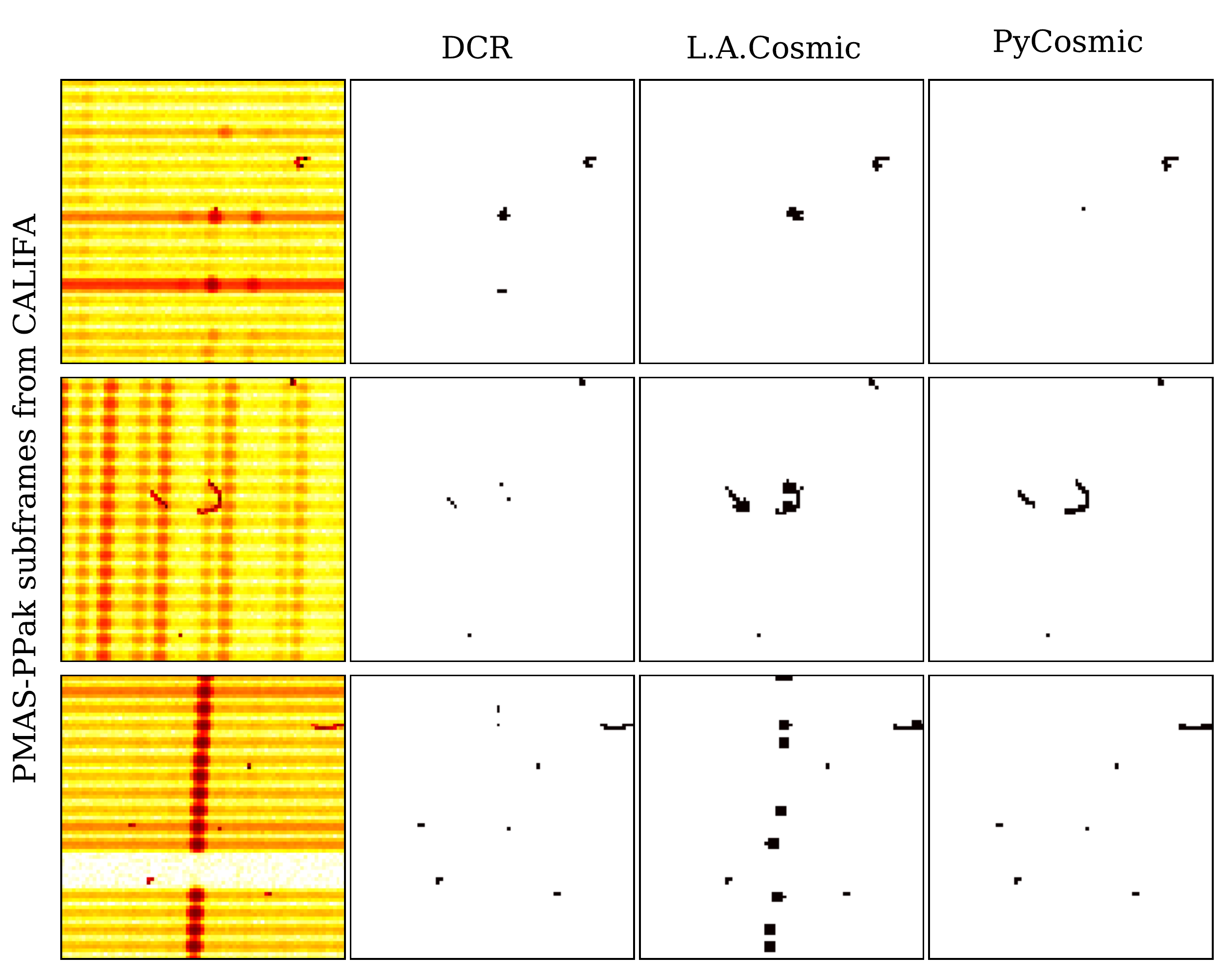}}\\
\resizebox{\hsize}{!}{\includegraphics{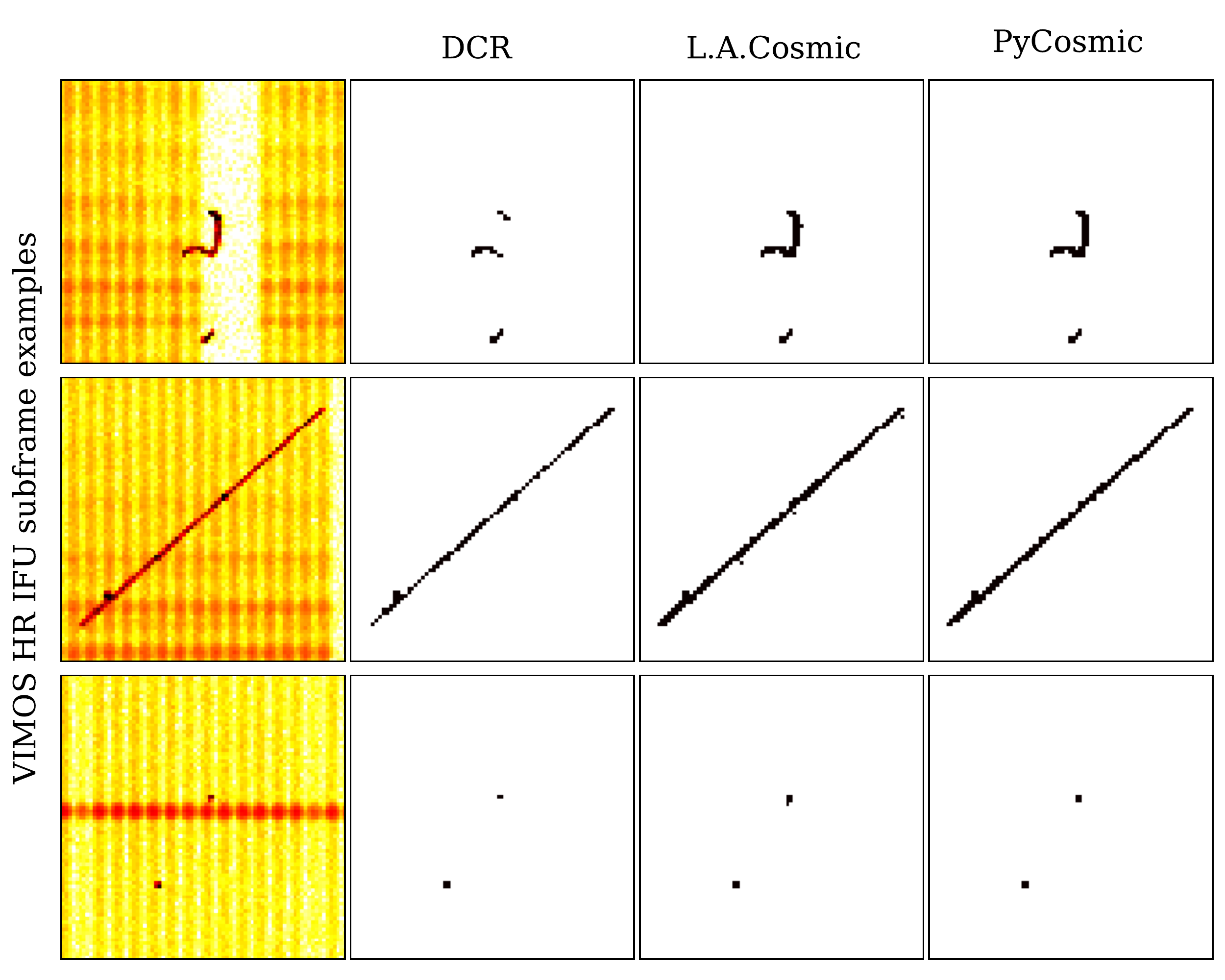}}
\caption{Comparison of cosmics detected in real observation taken with three different IFS instruments. Three representative subframes of the raw images (left column of thumbnail images) were chosen to allow a good comparison for the results of the \texttt{DCR}, the \texttt{L.A.Cosmic}, and the \texttt{PyCosmic} algorithms. Pixel masks of detected cosmics are shown in the three right panel columns.}
\label{fig:real}
\end{figure}

Although our simulations closely matched real observations, it is difficult to simulate realistic data, including object signal, given the complexity of IFS data. Thus, we similarly processed data from real observations for the different IFUs to check whether results from the simulations agreed well with those of observed data. The optimal parameters as inferred from the simulations are used for the different algorithms and are applied to a typical raw frame from the CALIFA survey taken with the PMAS-PPak IFU at 900\,s exposure time, to 1100\,s frame of a low-redshift galaxy taken with the VIMOS IFU in MR mode, and to a 1800\,s frame exposure of the center of a globular cluster taken with the PMAS-LArr IFU using the R1200(bw) setup. Representative subframes and corresponding cosmic masks recovered by the algorithms are shown in Fig.~\ref{fig:real} for the different datasets.

In general, results obtained for real data reflected the outcome of the mock data analysis. Cosmics of the selected subframes are illustrative of the strengths and weaknesses of the algorithms. In case of PMAS-PPak data, we  clearly see that \texttt{DCR} had problems detecting cosmics if the underlying signal was already quite high, yet it had few false detections at the same time. The false detections were a huge problem for \texttt{L.A.Cosmic}, not only in the simulations, because bright emission lines in real data were also classified as cosmics. \texttt{PyCosmic} almost perfectly detected cosmics and was by far the best algorithm for the PMAS-PPak instrument. This confirmed the necessity to develop a new algorithm for the CALIFA survey.

\section{Guidelines for algorithm selection and optimal parameter settings}\label{sect:guide}
Based on the performance of the individual algorithms on simulated and real data, we try to provide useful guidelines here for users that need to tackle the problem of cosmics detection during the reduction of IFS data.
\subsection{\texttt{DCR}}
The performance of the algorithm depends only weakly on the subframe size. We recommend a symmetric size of the order of $15\times15$ pixels. The main parameter to be set properly is $\xi$, which should be  $\xi\sim3$ to achieve the highest performance. An exception to these recommendations are undersampled data, where the subframe size seems to be important. A much higher value of $\xi$ needs to be chosen in this case ($\xi\geq8$) to achieve an acceptable false detection rate (see Fig.~\ref{fig:parameter3}). Because of the intrinsically lower detection rate compared to the other two algorithms, DCR should only be used in case the highest computational speed outweighs all other concerns. 

\subsection{\texttt{L.A.Cosmic}}
\begin{figure*}
\centering
 \includegraphics[width=0.9\textwidth]{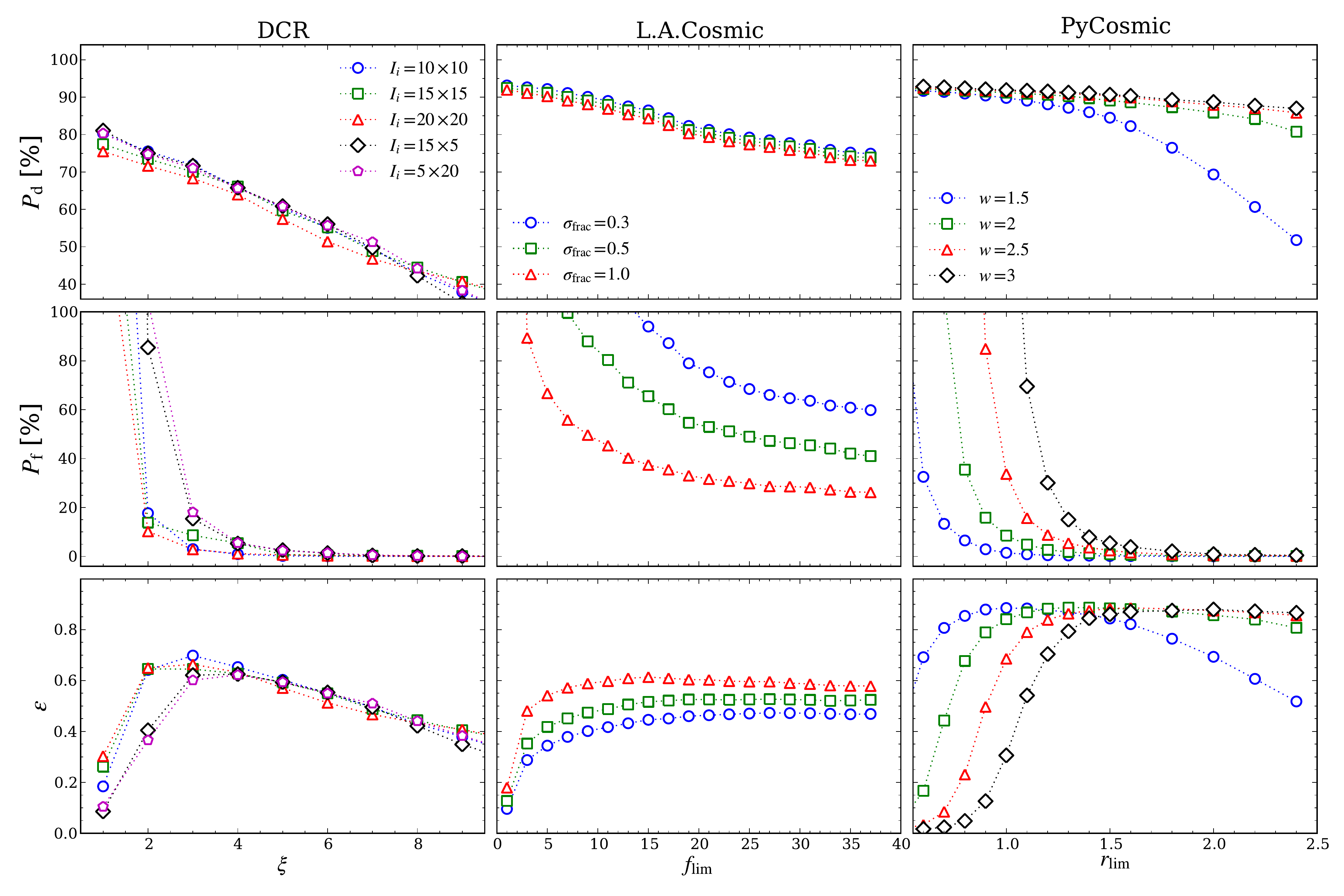}
  \caption{Same as Fig.~\ref{fig:parameter1} for simulated PMAS-PPak data, $\theta=2.4$\,pixels (CALIFA survey type data).}
 \label{fig:parameter2}
\end{figure*}
\begin{figure*}
\centering
 \includegraphics[width=0.9\textwidth]{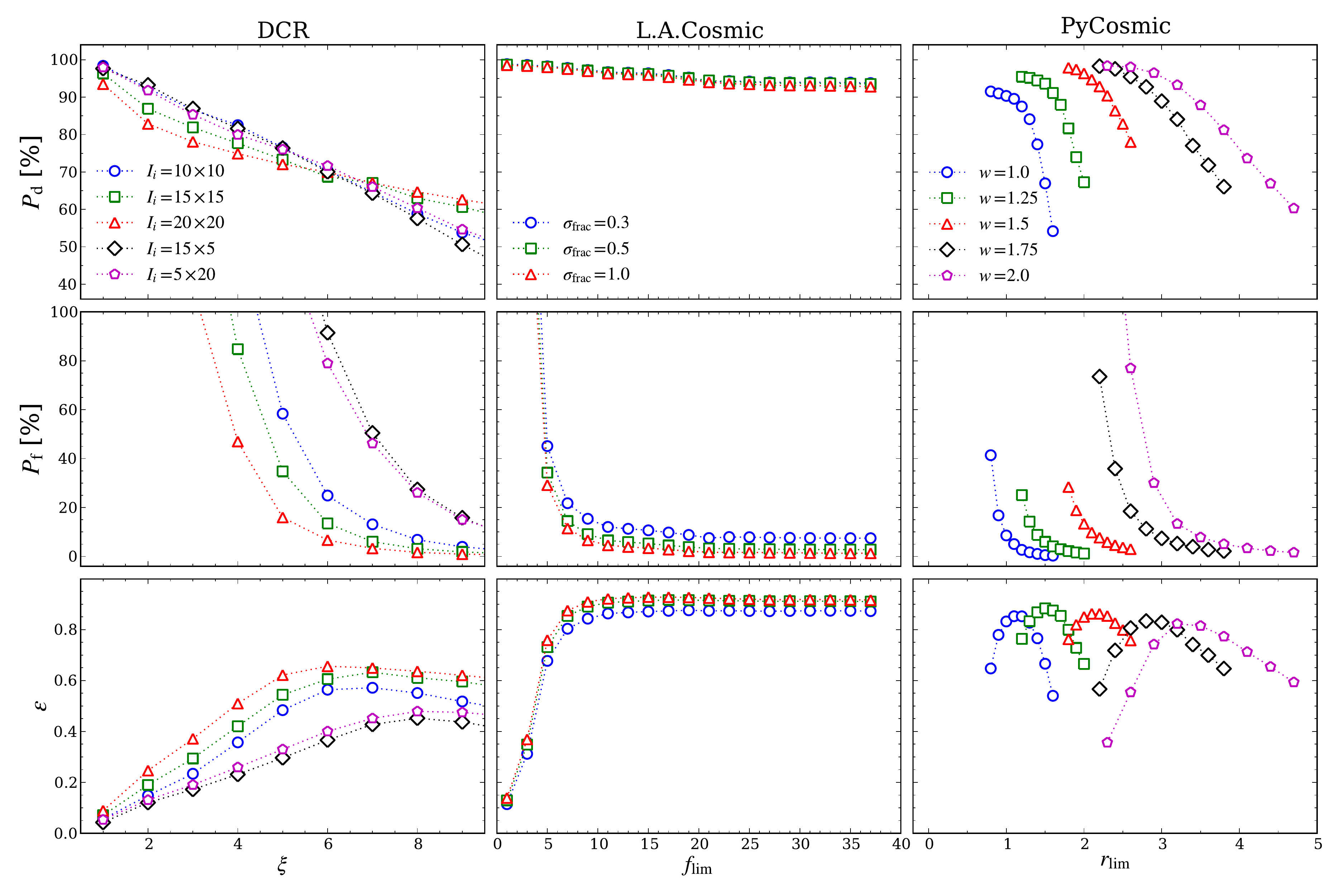}
  \caption{Same as Fig.~\ref{fig:parameter1} for the simulated PMAS-LArr data that are heavily undersampled, with $\theta=1.5$\,pixels.}
 \label{fig:parameter3}
\end{figure*}
The results of \texttt{L.A.Cosmic} are mildly dependent on the growing radius. The middle columns of Figs.~\ref{fig:parameter1}--\ref{fig:parameter3} show the best results using a zero growing radius. In this case, $\sigma_\mathrm{frac}\!=\!1.0$, which is our recommended value. The improvement that can be achieved with this parameter is relatively small when compared to models using a growing radius of one pixel and $\sigma_\mathrm{frac}\!=\!0.5$. %
Additionally, $P_\mathrm{d}$ and $P_\mathrm{f}$ vary smoothly with $f_\mathrm{lim}$ above a certain limit, but it is difficult to predict an optimal value of $f_\mathrm{lim}$ for a given instrument. In general, a value of $f_\mathrm{lim}>5$ is required even for well-sampled data. This is in contrast to the behavior and guidelines for imaging data given by D01. While \texttt{L.A.Cosmic} is able to reach an excellent performance for some IFS data, it is not a robust algorithm because it fails to produce acceptable results for certain IFU instrument configurations (see Fig.~\ref{fig:parameter2}).  \texttt{L.A.Cosmic} may be appropriate for undersampled IFS data, but it should not be applied to other IFS datasets without careful checking of the results.

\subsection{\texttt{PyCosmic}}
The width of the smoothing kernel should be set to $w\lesssim\theta$; otherwise, the optimal performance of \texttt{PyCosmic} cannot be reached. This is most evident in  undersampled data, where the maximum efficiency decreases substantially with increasing $w$ (Fig.~\ref{fig:parameter3}). For any given combination of $w$ and $\theta$, $r_\mathrm{lim}$ determines the efficiency of the detection. In extreme cases, as with undersampled data, the tolerance in $r_\mathrm{lim}$ to reach the optimal efficiency is small. However, comparing the theoretically derived minimum values of $r_\mathrm{lim}$ (cf. Fig.~\ref{fig:flim}) with the best values of the simulation, we consistently found  that the optimal $r_\mathrm{lim}$ threshold needs to be a factor of $\sim$2 larger than estimated from Fig.~\ref{fig:flim}. With these parameter settings, \texttt{PyCosmic} provides the most robust detection efficiency for any IFS instrument configuration with an efficiency of $\epsilon\gtrsim90$\%, well-defined parameter settings, and the possibility of reducing the number of false detections $P_\mathrm{f}$ to nearly zero.

\subsection{Handling cosmics in IFS data reduction}
All algorithms attempt to restore the information of pixels that are affected by cosmics. Nevertheless, the restored signal should be considered unreliable given the signal structure in IFS data. Instead of the common practice of simply processing the ``cleaned'' image, we emphasize that bad pixels can be nicely handled during the spectra extraction process when an optimal extraction scheme is used \citep[e.g.,][]{Horne:1986,Sharp:2010}. Given that optimal extraction algorithms always assume a certain shape of the signal on the CCD, it is easy to mask bad pixels and restore the signal at the cost of a higher associated variance. When too many pixels are affected by cosmics on the raw frame for a given spectral-resolution element, they should be flagged as bad elements that are propagated through the reduction pipeline to the final data product. We consider this to be the best possible scheme to handle artefacts caused by cosmics  in IFS data.

\section{Conclusions}
In this paper we have presented a novel detection algorithm for cosmics in single exposures called \texttt{PyCosmic}. The algorithm combines Laplacian edge detection with a PSF convolution approach. We systematically compared the performance of our new algorithm  against other standard detection algorithms, \texttt{DCR} \citep{Pych:2004} and \texttt{L.A.Cosmic} \citep{Dokkum:2001}, on simulated and real images from fiber-fed IFS instruments.  With the aid of these detailed comparison tests, we provide general recommendations for the use of these algorithms for the detection of cosmics in IFS data.

We have found that \texttt{DCR} does not reach a detection efficiency equivalent to that of \texttt{L.A.Cosmic} and \texttt{PyCosmic}. Therefore, we cannot recommend its use for IFS data in general, except when computational speed is critical. The strength of the \texttt{L.A.Cosmic} algorithm is that it works best for undersampled IFS data. However, a significant drawback is that the minimum false detection rate achievable for a given IFS data is entirely set by the characteristics of the instrument and cannot be reduced by changing any parameter settings. This peculiarity is most evident for PMAS-PPak IFU data from the CALIFA survey \citep{Sanchez:2012a}, where the false detection rate of \texttt{L.A.Cosmic} is $P_\mathrm{f}\gtrsim40\%$.  Our \texttt{PyCosmic} algorithm reduces the false detection rate with different parameter settings and solves this problem effectively. It has replaced the simplified \texttt{R3D} routine (based solely on a Laplacian edge detection scheme) in the reduction pipeline of the CALIFA survey.

\texttt{PyCosmic} is the most robust detection algorithm for cosmics in fiber-fed IFS data. In combination with well-characterized optimal parameter settings, it is well-suited for automatic usage for very large datasets. CALIFA is already a huge IFS survey by current standards that has significantly benefited from the development of \texttt{PyCosmic}. The next generation of IFS instruments like the Sydney-AAO Multi-object IFS \citep[SAMI,][]{Croom:2012} or the IFU project Mapping Nearby Galaxies at APO (MaNGA) is already being built or is planned to carry out even larger IFS surveys. These surveys will deliver IFS data for thousands of galaxies in the near future, which will certainly benefit from robust data reduction algorithms such as \texttt{PyCosmic}.  

The \texttt{PyCosmic} algorithm has recently been implemented in the versatile multi-IFU reduction software \texttt{P3D} \citep{Sandin:2010} and is also available as a \texttt{Python}-based stand-alone program\footnote{\texttt{PyCosmic} is available for download at \url{http://pycosmic.sf.net}} so that it can be easily used or even added to any existing IFS reduction pipeline. Although \texttt{PyCosmic} has been optimized for IFS data, we have also applied it successfully to longslit data and anticipate that good results will be achieved with imaging data.

\begin{acknowledgements} 
We thank the anonymous referee for a very prompt report with instructive comments that improved the quality and presentation of the article.
Furthermore, we thank Ana Monreal Ibero for valuable comments and Barry Rothberg for several suggestions that increased the readability of the manuscript.
B.~H. gratefully acknowledge the support by the DFG via grant Wi 1369/29-1.
C.~S.\ was supported by the grant PTDESY-05A12BA1, and R.~G.-B. acknowledges support from the Spanish Ministerio de Ciencia e Innovacion through grant AYA2010-15081.
\end{acknowledgements}   

\bibliographystyle{aa}
\bibliography{references}

\end{document}